\documentstyle[prl,aps]{revtex}
\input epsf

\begin{document}

\title{Constrained orthogonal polynomials}

\author{B.G. Giraud,}
\address{giraud@spht.saclay.cea.fr, Service de 
Physique Th\'eorique, DSM, CE Saclay, F-91191 Gif/Yvette, France}

\date{\today}
\maketitle

\begin{abstract}

Abstract: We define sets of orthogonal polynomials satisfying the additional
constraint of a vanishing average. These are of interest, for example, for 
the study of the Hohenberg-Kohn functional for electronic or nucleonic 
densities and for the study of density fluctuations in centrifuges. We give 
explicit properties of such polynomial sets, generalizing Laguerre and 
Legendre polynomials. The nature of the dimension 1 subspace completing such 
sets is described. A numerical example illustrates the use of such polynomials.

\end{abstract}

\section{Introduction}

Generalizations $\Gamma_n$ of Hermite polynomials $H_n$ were recently 
\cite{GiMeWe} proposed to describe, for instance, density perturbations 
constrained by a condition of matter conservation. Because of the constraint, 
such polynomials cannot form a complete set, but span a subspace well suited 
to specific applications. In particular, the polynomials $\Gamma_n$ used in 
\cite{GiMeWe} were motivated by the consideration in nuclear physics of the 
Hohenberg-Kohn functional \cite{HK} and similar functionals along the 
Thomas-Fermi method \cite{Mer,WKS}. Indeed, in such approaches, the ground 
state of a quantum system is shown to be a functional of its density 
$\rho(r),$ and there is a special connection between $\rho(r)$ and the mean 
field $u(r)$ driving the system. It was thus convenient to expand variations 
of $\rho$ in a basis $\{w_m(r)\}$ of particle number conserving 
components, $\delta \rho(r)=\sum_m\, \delta \rho_m\, w_m(r),$ with the 
term-by-term constraint, $\forall m,\ \int dr\, w_m(r)=0.$ This spares, in the 
formalism, the often cumbersome use of a Lagrange multiplier. Simultaneously, 
it was convenient to expand variations of $u$ in a basis orthogonal to the 
flat potential, because, trivially, a flat $\delta u,$ as just a change in 
energy reference, cannot influence the density. The same basis can thus be 
used for $\delta u(r)=\sum_n \delta u_n\, w_n(r),$ since the very same 
condition, $\int dr\, w_n(r)=0,$ induces orthogonality to a constant 
$\delta u.$ Because of the nuclear physics context of \cite{GiMeWe}, 
harmonic oscillators shell models were considered and the basis contained 
a Gaussian factor, $e^{-\frac{1}{2}r^2}.$

The same functional approaches \cite{HK,Mer,WKS} are also of a general use
in atomic and molecular physics, where Gaussian weights would be clumsy and 
radial properties are best fitted with simple exponential weights 
\cite{Messiah}. Furthermore, in \cite{GiMeWe}, the discussion was restricted 
to one dimensional problems. In the present note, we want to include two and 
three dimensional situations. We shall thus use weights of the form 
$e^{-\frac{1}{2}r},$ with $0 \le r < \infty,$ but integrals will carry a 
factor $r^{\nu},$ with $\nu$ a positive exponent, suitable for dimension $d.$ 
This will lead to generalizations of Laguerre polynomials.

This note is also concerned with compact domains, of the form $0 \le r \le 1$ 
for instance. This might correspond for instance to expansions of density 
fluctuations in cylindrical vessels used for chemical processes, where mass 
conservation is also in order, or maybe in centrifuges. Radial integrals 
with factors $r$ and $r^2$ in both the constraint and orthogonalization
conditions will lead to generalizations of Legendre polynomials.

For any positive weight $\mu(r),$ and any dimension $d,$ a constraint of 
vanishing average, $\int dr\, r^{\nu}\, \mu(r)\, \Gamma_n(r) =0,$ is 
incompatible with a polynomial $\Gamma$ of order $n=0.$ Therefore, in the 
following, the order hierarchy for the constrained polynomials runs 
from $n=1$ to $\infty,$ while that for the traditional polynomials runs from 
$0$ to $\infty.$ We study in some generality the ``Laguerre'' case in 
Section II. In turn, the ``Legendre'' case is the subject of Section III. 
A brief Section IV discusses possible applications to the study of density 
fluctuations in centrifuges. Section V answers a question which was omitted 
in \cite{GiMeWe}, that of the nature of the projector onto the subspace 
spanned by the constrained states and the nature of the codimension of this 
subspace. A numerical application is provided in Section VI. A discussion and 
conclusion make Section VII.

\section{Modification of Laguerre polynomials by a constraint of zero average}

In this Section we consider basis states carrying a weight 
$e^{-\frac{1}{2}r},$ in the form $w_n(r)=e^{-\frac{1}{2}r}\, G_n^d(r),$
where $G_n^d$ is a polynomial. It is clear that $G_0^d$ cannot be a finite,
non vanishing constant if the constraint, 
$\int_0^{\infty}dr\, r^{d-1}\, e^{-\frac{1}{2}r}\, G_0^d(r)=0,$ must be 
implemented. Hence set integer labels $m \ge 1$ and $n \ge 1$ and define 
polynomials $G_n^d$ by the conditions,
\begin{equation}
\int_0^\infty dr\ r^{d-1}\, e^{-r}\, G_m^d(r)\, G_n^d(r) = g_n^d\, 
\delta_{mn}\, ,
\ \ \ \ \ 
\int_0^\infty dr\ r^{d-1}\, e^{-\frac{1}{2}r}\, G_n^d(r) = 0\, , 
\label{definL}
\end{equation}
where $\delta_{mn}$ is the usual Kronecker symbol and the positive numbers 
$g_n^d$ are normalizations, to be defined later.

It is elementary to generate such polynomials numerically, in two steps by 
brute force, namely i) first create ``trivial seeds'' of the form, 
$s_n^d(r)=r^n-\langle r^n \rangle_d,$ where the subtraction of the average, 
$\langle r^n \rangle_d=2^n\, (d-1+n)!/(d-1)!,$ ensures that each trivial 
seed fulfills the constraint, then ii) orthogonalize such seeds by a 
Gram-Schmidt algorithm. The first polynomials read,
\begin{mathletters} \begin{eqnarray}
G_1^1 = r-2,\ \ \ \, G_2^1 = r^2-5r+2,\ \ \ \ \ G_3^1 = r^3-10r^2+20r-8,\ \ 
\ \ \ \ \ G_4^1 = r^4-17r^3+78r^2-108r+24, \label{G1} \\
G_1^2 = r-4,\ \ \ \ G_2^2 = r^2-8r+8,\ \ \ \ G_3^2 = r^3-14r^2+44r-32,\ \ \ 
G_4^2 = r^4-22r^3+138r^2-288r+144, \label{G2} \\
G_1^3 = r-6,\ \ \  G_2^3 = r^2-11r+18,\ \ \  G_3^3 = r^3-18r^2+78r-84,\ \ \  
G_4^3 = r^4-27r^3+216r^2-606r+468. \label{G3} 
\end{eqnarray} \end{mathletters}
All these are defined to be ``monic'', namely the coefficient of $r^n$ is 
always $1.$ For an illustration we show in Figure 1 the new 
polynomials $G_1^1$ and $G_1^2,$ together with Laguerre polynomial $ L_1.$ 
The same Fig. 1 also shows $G_2^1,$ $G_2^2$ and $L_2.$

\begin{figure}[htb] \centering
\mbox{  \epsfysize=100mm
         \epsffile{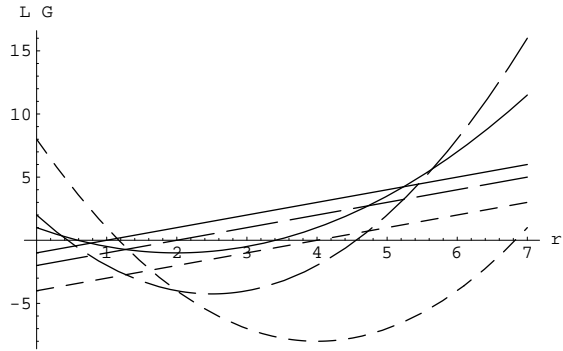}
     }
\caption{Comparison of Laguerre polynomials $L_1,$ $L_2$  (full lines)
with new polynomials $G_1^1,$  $G_2^1$ (long dashes), $G_1^2,$ $G_2^2$
(dashes).}
\end{figure}

Rather using the Gram-Schmidt method, we find it easier, and more elegant, to 
generate the polynomials $G_n^d,$ starting from the initial table, 
Eqs. (\ref{G1},\ref{G2},\ref{G3}), by means of the following recursion formula,
\begin{equation}
G_n^d(r) = (r-d)\, G_{n-1}^d(r)-2\, r\, G_{n-1}^{d\, \prime}(r)+(n+d-1)\, (n-2)
\, G_{n-2}^d(r),
\label{recurG}
\end{equation}
where the prime denotes the derivative with respect to $r.$ Its simple 
structure can be proven analytically as follows:

i) Let us first create some kind of a ``less trivial seed'' at order $n,$ 
assuming the polynomial $G_{n-1}^d$ is known. For this, try $r\, G_{n-1}^d.$ 
By partial integration, we see that,
\begin{equation}
\int_0^\infty dr\ r^{d-1}\, e^{-\frac{1}{2}r} \left[r\, G_{n-1}^d(r)\right] = 
 2 \int_0^\infty dr\ e^{-\frac{1}{2}r} \left[r^d\, G_{n-1}^d(r)\right]' \, ,
\label{seedG}
\end{equation}
where again a prime means derivation with respect to $r.$ Thus 
$\sigma_n^d \equiv \left(r\, G_{n-1}^d - 2\, r\, G_{n-1}^{d\, \prime} - 2\, 
d\, G_{n-1}^d \right)$ makes indeed a less trivial seed, compatible with 
the constraint. Notice that the order $n$ of this seed polynomial $\sigma_n^d$ 
comes from the term $r\, G_{n-1}^d$ only, the other two terms having 
order $n-1.$ Notice again that, in the table, Eqs. (2), all polynomials 
$G_n^d$ are monic. We can define $G_n^d$ as monic, systematically. Since 
the product $r\, G_{n-1}^d$ respect this ``monicity'', and since $\sigma_n^d$ 
fulfills the constraint, we conclude that $\sigma_n^d$ is a linear 
combination of $G_n^d,$ with coefficient 1, and of all the lower order 
polynomials $G_m^d,$ with $1\le m<n,$ but with yet unknown coefficients.

ii) It turns out that such coefficients vanish if $m < n-2.$ Indeed, an 
integration of  $\sigma_n^d$ against $G_m^d,$ weighted by $r^{d-1} e^{-r},$
gives, by partial integration of the $G_{n-1}^{d\, \prime}$ term, 
\begin{eqnarray}
&\int_0^{\infty} dr\ e^{-r}\, r^{d-1}\, \sigma_n^d(r)\, G_m^d(r) \equiv
\int_0^{\infty} dr\ r^{d-1} e^{-r} 
\left[ (r-2d)\, G_{n-1}^d(r) - 2r\, G_{n-1}^{d\, \prime}(r) \right]
\, G_m^d(r) = \nonumber \\
&\int_0^{\infty} dr\ e^{-r}\, r^{d-1}\, G_{n-1}^d(r)\, (r-2d)\, G_m^d(r)
+ 2\int_0^{\infty} dr\ G_{n-1}^d(r)\, \left[ e^{-r}\, r^d\, G_m^d(r)\right]' =
\nonumber \\
&\int_0^{\infty} dr\ e^{-r}\, r^{d-1}\, G_{n-1}^d(r) 
\left[-\sigma_{m+1}^d(r)-2d\, G_m^d(r)\right] .
\label{simplifG}
\end{eqnarray}
In the bracket $[\ ]$ in the last right-hand side of Eq. (\ref{simplifG}) the 
seed $\sigma_{m+1}^d$ has order $m+1$ and, by definition, $G_m^d$ is of order 
$m.$ By definition also, $G_{n-1}^d,$ of order $n-1,$ is orthogonal to all 
those polynomials of lower order, that are compatible with the constraint. 
This integral, Eq. (\ref{simplifG}), thus vanishes as long as $m+1 < n-1.$ It 
can be concluded that the difference, $\sigma_n^d-G_n^d,$ contains only two 
contributions, namely those from $G_{n-2}^d$ and $G_{n-1}^d.$ Explicit forms 
for their coefficients are obtained by elementary manipulations, leading to 
Eq. (\ref{recurG}).
Elementary manipulations also give, 
\begin{equation}
2\, r\, G_n^{d\, \prime \prime} - (r-2d)\, G_n^{d\, \prime} + n\, G_n^d = 
(n-1)\, (n+d)\, G_{n-1}^d\, .
\end{equation}
Here, in the same way as a prime means first derivative with respect to $r,$ 
we used double primes for second derivatives. 
Finally the normalization of the polynomials is obtained easily as,
\begin{equation}
g_n^d \equiv \int_0^{\infty} dr\ e^{-r}\, r^{d-1}\, [G_n^d(r)]^2 = 
(n-1)!\, (n+d)!\, .
\end{equation}

\section{Modification of Legendre polynomials by a constraint of zero average}

Legendre polynomials, and their associates and generalizations 
(Gegenbauer, Chebyshev, Jacobi) are defined with respect to the $[-1,1]$ 
segment. Exceptionally in the literature, one finds shifted Legendre 
polynomials, adjusted to the $[0,1]$ segment. We are here interested in 
applications to radial densities in cylinders, or the small circles of 
toruses, or spheres. Hence we shall use $0 \le r \le 1.$ It is clear that the 
case, $d=1,$ does not make an original problem, since Legendre polynomials, 
whether translated and/or scaled or not, already average to $0$ as soon as 
their order $n$ is $\ge 1.$  We keep the case, $d=1,$ for the sake only of 
completeness and in this Section we consider $d=1,2,3,$ with a geometry 
factor $r^{d-1}.$ The weight is $\mu(r)=1,$ hence our states are described 
by just a polynomial ${\cal G}_n^d$ of order $n.$ It is again obvious that 
${\cal G}_0^d$ cannot be a non vanishing constant if the constraint, 
$\int_0^1 dr\, r^{d-1}\, {\cal G}_0^d(r)=0,$ is implemented. Hence set 
$m \ge 1,$ $n \ge 1,$ and define polynomials ${\cal G}_n^d$ from conditions,
\begin{equation}
\int_0^1 dr\ r^{d-1}\, {\cal G}_m^d(r)\, {\cal G}_n^d(r) = \gamma_n^d\, 
\delta_{mn}\, ,
\ \ \ \ \ 
\int_0^1 dr\ r^{d-1}\, {\cal G}_n^d(r) = 0
\, , 
\label{definl}
\end{equation}
where the normalizations $\gamma_n^d$ are again to be defined later. 
It is obvious that the shifted (and shrunk) Legendre polynomials 
${\cal L}_n(2r-1),\ n \ge 1,$ satisfy both constraint and  orthogonality 
relations for $d=1,$ because they are orthogonal to any constant polynomial, 
of order $0.$ The polynomials ${\cal G}_n^1={\cal L}_n(2r-1)$ thus make  
nothing new. We turn therefore to $d=2$ and $d=3,$ with a brute force 
construction as in the previous Section. But the defining conditions, 
Eqs. (\ref{definl}), show a difference with Eqs. (\ref{definL}): both 
orthogonality and constraint conditions now carry the same weight, 
namely $\mu^2=\mu,$ while in the previous case, Eqs. (\ref{definL}), there 
were different weights, because of the exponentials $e^{-r}$ and 
$e^{-\frac{1}{2}r}.$ A similar difference between $\mu^2$ and $\mu$ happened 
in the ``Hermite'' case, naturally. 

Hence now, in this Legendre case, we can  Gram-Schmidt orthogonalize 
even more trivial seeds $r^n,$ without subtractions, and accept those 
orthogonal polynomials with order $m \ge 1.$ The table of first results reads,
\begin{mathletters} \begin{eqnarray}
{\cal G}_1^1 = 2r-1,\ \ \ \ \, {\cal G}_2^1 = 6r^2-6r+1,\ \ \ \ 
{\cal G}_3^1 = (2r-1) (10 r^2-10r+1),\ \ \ \
{\cal G}_4^1 = 70r^4-140r^3+90r^2-20r+1, \label{Gl1} 
\\
{\cal G}_1^2 = 3r-2,\ \, {\cal G}_2^2 = 10r^2-12r+3,\ \ \ \ \, 
{\cal G}_3^2 = 35r^3-60r^2+30r-4,\ \ \ \ \,
{\cal G}_4^2 = 126r^4-280r^3+210r^2-60r+5, \label{Gl2} 
\\
{\cal G}_1^3 = 4r-3,\ {\cal G}_2^3 = 15r^2-20r+6,\ \  
{\cal G}_3^3 = 56r^3-105r^2+60r-10,\ \  
{\cal G}_4^3 = 210r^4-504r^3+420r^2-140r+15. \label{Gl3} 
\end{eqnarray} \end{mathletters}

Easy, but slightly tedious manipulations validate the following recursion 
relations,
\begin{mathletters} \begin{eqnarray}
n\, {\cal G}_n^1 &=& 
(2n-1)\, (2r-1)\, {\cal G}_{n-1}^1 - (n-1)\, {\cal G}_{n-2}^1\, , \\
(n+1)\, (2n-1)\, {\cal G}_n^2 &=&
2\, [\,(4n^2-1)r-2n^2\,]\, {\cal G}_{n-1}^2 - (n-1)\, (2n+1)\, 
{\cal G}_{n-2}^2\, , \\
n^2\, (n+2)\, {\cal G}_n^3 &=&
(2n+1)\, [\, 2n(n+1)r - (n^2+n+1)\, ]\, {\cal G}_{n-1}^3 - (n-1)\, (n+1)^2\,
\, {\cal G}_{n-2}^3\, .
\end{eqnarray} \end{mathletters}
and the differential equation,
\begin{equation}
r\, (r-1)\, {\cal G}_n^{d\, \prime \prime} + [\, (d+1)\, r-d\,]\, 
{\cal G}_n^{d\, \prime} - n\, (n+d)\, {\cal G}_n^d = 0.
\end{equation}

Finally the normalization of the polynomials reads,
\begin{equation}
\gamma_n^d \equiv \int_0^{\infty} dr\ r^{d-1}\, [{\cal G}_n^d(r)]^2 = 
1/(2n+d)\, .
\end{equation}
We show in Figure 2 the plots of ${\cal G}_n^d$ for $n=1,2$ and $d=1,2,3.$

\begin{figure}[htb] \centering
\mbox{  \epsfysize=100mm
         \epsffile{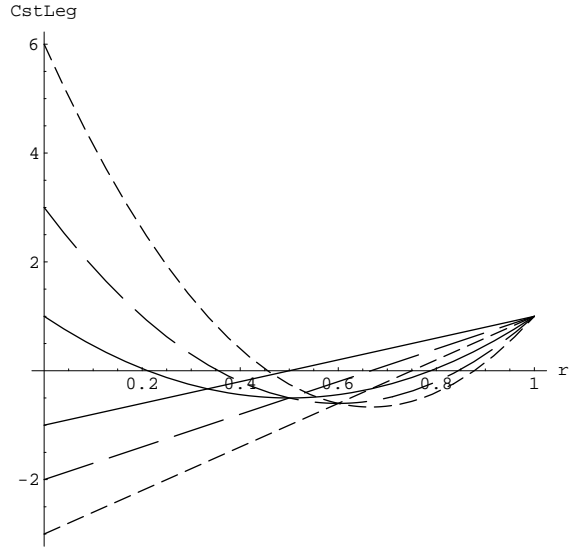}
     }
\caption{Modified Legendre polynomials ${\cal G}_1^1,$ ${\cal G}_2^1$  
(full lines), ${\cal G}_1^2,$  ${\cal G}_2^2$ (long dashes), ${\cal G}_1^3,$ 
${\cal G}_2^3$ (dashes).}
\end{figure}

\section{Polynomials for centrifuges}

The case of centrifuges is worth a short comment. As soon as the matter under 
centrifugation is compressible, the density becomes much larger at the outer 
edge, $r=1,$ than at the rotation axis, $r=0.$ Let $h$ be the height
of the centrifuge. Assume, for the sake for the argument, that one studies 
fluctuations about a reference density of the form, 
$\rho(r)=\rho_c\, e^{Kr^2},$ where the parameter $K$ contains 
all informations about the angular velocity, compressibility, etc. of the 
process. The factor,
$\rho_c = M\, \left[h\, \int_0^1 dr\, r\, \rho(r)\right]^{-1} =
M\, h^{-1}\, 2\, K\, \left[e^K-1\right]^{-1},$ 
ensures the conservation of the mass $M$ included in the vessel. If a cause 
for fluctuations of $\rho$ is an instability of $K,$ the first order for 
density change is,
\begin{equation}
\frac{\partial \rho}{\partial K}(r)=2\, \frac{K\, r^2\, 
\left[e^K-1\right]+ e^K - K e^K -1}{\left[e^K-1\right]^2}\, e^{Kr^2}\, , \ \ \
\ \ \ \int_0^1 dr\, r\, \frac{\partial \rho}{\partial K}(r)=0,
\end{equation}
namely a polynomial of order 2 multiplied by $e^{Kr^2}.$ Higher derivatives 
with respect to $K$ will generate similar, even order polynomials, with the 
same property, $\int_0^1 dr\, r\, \partial^n \rho/ \partial K^n(r)=0.$ An
orthogonalization, under a metric $\propto e^{2Kr^2},$ might be useful. 
This new set of polynomials will depend on $K,$ however, since $r$ is already 
scaled to a radius $1$ for the cylinder and thus $K$ cannot be scaled away.
Because of this $K$ dependence we do not elaborate further on this issue.
For a large list of {\it ad hoc} polynomials and integration weights, see 
\cite{JPB}.

\section{Projector on the constrained subspace}

For the sake of the discussion and short notations, set first $d=1,$
$\mu(r)=e^{-\frac{1}{2}r},$ and temporarily include normalization to unity
factors into both the Laguerre polynomials $L_n$ and the constrained 
$G_n^1.$ This summarizes as, 
\begin{equation}
\int_0^{\infty} dr\ [\mu(r)]^2\, L_m(r)\, L_n(r)=\delta_{mn},\ \ \ \ \ 
\int_0^{\infty} dr\ [\mu(r)]^2\, G_m^1(r)\, G_n^1(r)=\delta_{mn},
\ \ \ \ \ 
\int_0^{\infty} dr\ \mu(r)\, G_n^1(r)=0,
\label{scheme}
\end{equation}
Then the kets and bras defined by 
$\langle r | w_n \rangle = \langle w_n | r \rangle = w_n(r) = 
\mu(r)\, G_n^1(r)$ and 
$\langle r | z_n \rangle = \langle z_n | r \rangle = z_n(r) =
\mu(r)\, L_n(r)$ provide two ``truncation'' projectors, 
${\cal P}_N = \sum_{n=1}^N | w_n \rangle \langle w_n | $ and
${\cal Q}_N = \sum_{n=0}^N | z_n \rangle \langle z_n | ,$ available
for subspaces where polynomial orders do not exceed $N.$ Their respective 
ranks $N$ and $N+1,$ and the embedding and commutation relation,
$\left[{\cal P}_N,{\cal Q}_N\right]={\cal P}_N,$ are obvious. Obvious also is
the limit, $\lim_{N \rightarrow \infty} {\cal Q}_N =1.$ The role of the rank
one $| \sigma_N \rangle \langle \sigma_N | $ difference 
${\cal P}_N - {\cal Q}_N$ is to subtract from any test state, 
$| \tau \rangle = \sum_{n=0}^N \tau_n | z_n \rangle,$
that part which violates the condition of vanishing average. We shall show 
that the elementary ansatz,
\begin{equation}
| \sigma_N \rangle = \left( \sum_{m=0}^N\, \langle z_m \rangle^2 
\right)^{-\frac{1}{2}}\,
\sum_{n=0}^N\, \langle z_n \rangle\ | z_n \rangle,\ \ \ \ \ \ 
\langle z_n \rangle = \int_0^{\infty} dr\, \langle r | z_n \rangle,
\label{ansatz}
\end{equation}
defines the proper ``subtractor'' operator 
$| \sigma_N \rangle \langle \sigma_N |.$ Indeed, from
\begin{equation}
\left(\, {\cal Q}_N - | \sigma_N \rangle \langle \sigma_N |\, \right)\, 
| \tau \rangle = \sum_{n=0}^N \tau_n\, | z_n \rangle - 
\left( \sum_{m=0}^N \langle z_m \rangle^2 \right)^{-1}\, 
\left(\, \sum_{n=0}^N \langle z_n \rangle\, | z_n \rangle\, \right)\, 
\left( \sum_{p=0}^N \langle z_p \rangle\, \tau_p \right),
\end{equation}
one obtains
\begin{equation}
\int_0^{\infty} dr\, \langle r |\left( {\cal Q}_N - 
| \sigma_N \rangle \langle \sigma_N | \right)\, | \tau \rangle =
\sum_{n=0}^N \tau_n\, \langle z_n \rangle - 
\left( \sum_{m=0}^N \langle z_m \rangle^2 \right)^{-1}\,
\left(\, \sum_{n=0}^N \langle z_n \rangle\, \langle z_n \rangle\, \right)\, 
\left( \sum_{p=0}^N \langle z_p \rangle\, \tau_p \right)=0.
\end{equation}
Hence ${\cal Q}_N - | \sigma_N \rangle \langle \sigma_N |$ is the projector
${\cal P}_N.$ Incidentally, the Laguerre result for $\sigma_N$ is very simple,
because $\langle z_m \rangle = 2,\ \forall m.$ But the ansatz for $\sigma_N,$ 
Eq.(\ref{ansatz}), generalizes to all cases. For instance, with Hermite 
polynomials, odd orders already satisfy the constraint when integrated 
from $-\infty$ to $\infty,$ naturally, and thus do not 
contribute to $\sigma_N.$ Even orders contribute, and it is easy to verify,
upon integrating from $-\infty$ to $\infty$ again, that 
$\langle z_{2p} \rangle^2 = \pi^{\frac{1}{2}}\, 2^{1-p}\, (2p-1)!!/p!$.

It may be pointed out that the condition, $\int dr\, \mu(r)\, f(r)=0,$ for 
functions $f$ orthogonalized, like our polynomials, by a metric  
$[\mu(r)]^2,$ might be interpreted as an orthogonality condition, 
$\int dr\, f(r)\, [\mu(r)]^2\, g(r)=0,$ with $g(r)=[\mu(r)]^{-1}.$ This makes
$g$ a candidate for the subtractor form factor $\sigma.$ This
is of some interest for the centrifuge case, where a state function such as,
for instance, $e^{-Kr^2},$ remains finite when $0\le r \le 1.$ But there 
is little need to stress that, when the support of $\mu$ extends to $\infty,$ 
then $\mu^{-1}$ does not belong to the Hilbert space and cannot be used for
$\sigma.$ 

More interesting is the limiting process, $N \rightarrow \infty,$ as 
illustrated by Figures 2-5. Figs. 2 and 3 show the shapes, in terms of $r,$ 
of $\langle 2  | {\cal P}_N | r \rangle$ and 
   $\langle 10 | {\cal P}_N | r \rangle,$ respectively, when the projectors 
are made of the modified Laguerre polynomials $G_n^1.$ The build up of an 
approximate $\delta$-function when $N$ increases from $N=50$ (short dashes) 
to $N=100$ (long ones) and $N=150$ (full lines) is transparent, although the 
convergence is faster when peaks are closer to the origin, compare Figs. 2 
and 3. The slower convergence in Fig. 3 is due to the cut-off 
imposed by exponential weights as long as $N$ is finite. Given $N,$ there is 
a ``box effect'', the range of the box being of order $\sim N.$ A similar 
build up is observed for our other families of constrained polynomials, with 
slightly different details of minor importance such as, for instance, a box 
range $\sim \sqrt N$ for the Hermite case. 

The box effect is even more transparent in Figs. 4 and 5, which show the 
shapes of subtractors 
$\langle 10 | \sigma_N \rangle \langle \sigma_N | r \rangle$ and
$\langle  0 | \sigma_N \rangle \langle \sigma_N | r \rangle$ deduced 
from constrained polynomials of the Laguerre (Fig. 4) and Hermite (Fig. 5) 
type, respectively. (For graphical convenience, the polynomials 
$\Gamma_n^1$ and $H_n$ used for the Hermite case, Fig. 5, are tuned to 
a weight $e^{-r^2}$ rather than $e^{-\frac{1}{2}r^2},$ but this detail is 
not critical.)

It seems safe to predict that, given an effective length $\Lambda(N)$ for the 
box, the wiggles of the subtractor will smooth out when $N \rightarrow \infty$
and that only a background $\sim -1/\Lambda(N)$ will then remain.

\begin{figure}[htb] \centering
\mbox{  \epsfysize=100mm
         \epsffile{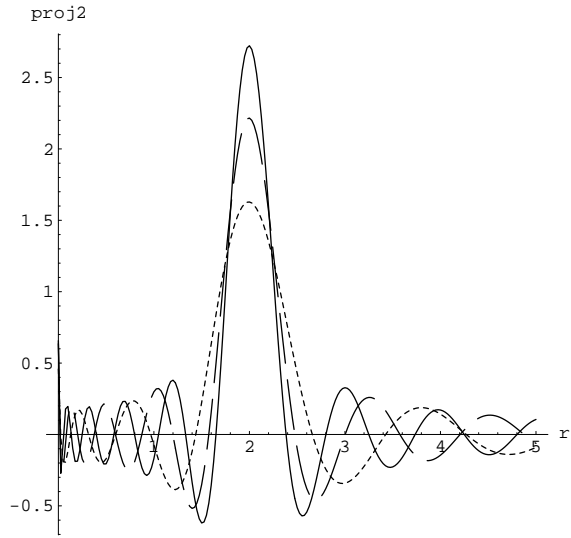}
     }
\caption{Shapes of projectors made of polynomials $G_n^1.$ Full line,
$\langle 2|{\cal P}_{150}|r \rangle,$ long dashes, 
$\langle 2|{\cal P}_{100}|r \rangle,$ short dashes
$\langle 2|{\cal P}_{50}|r \rangle.$}
\end{figure}

\begin{figure}[htb] \centering
\mbox{  \epsfysize=100mm
         \epsffile{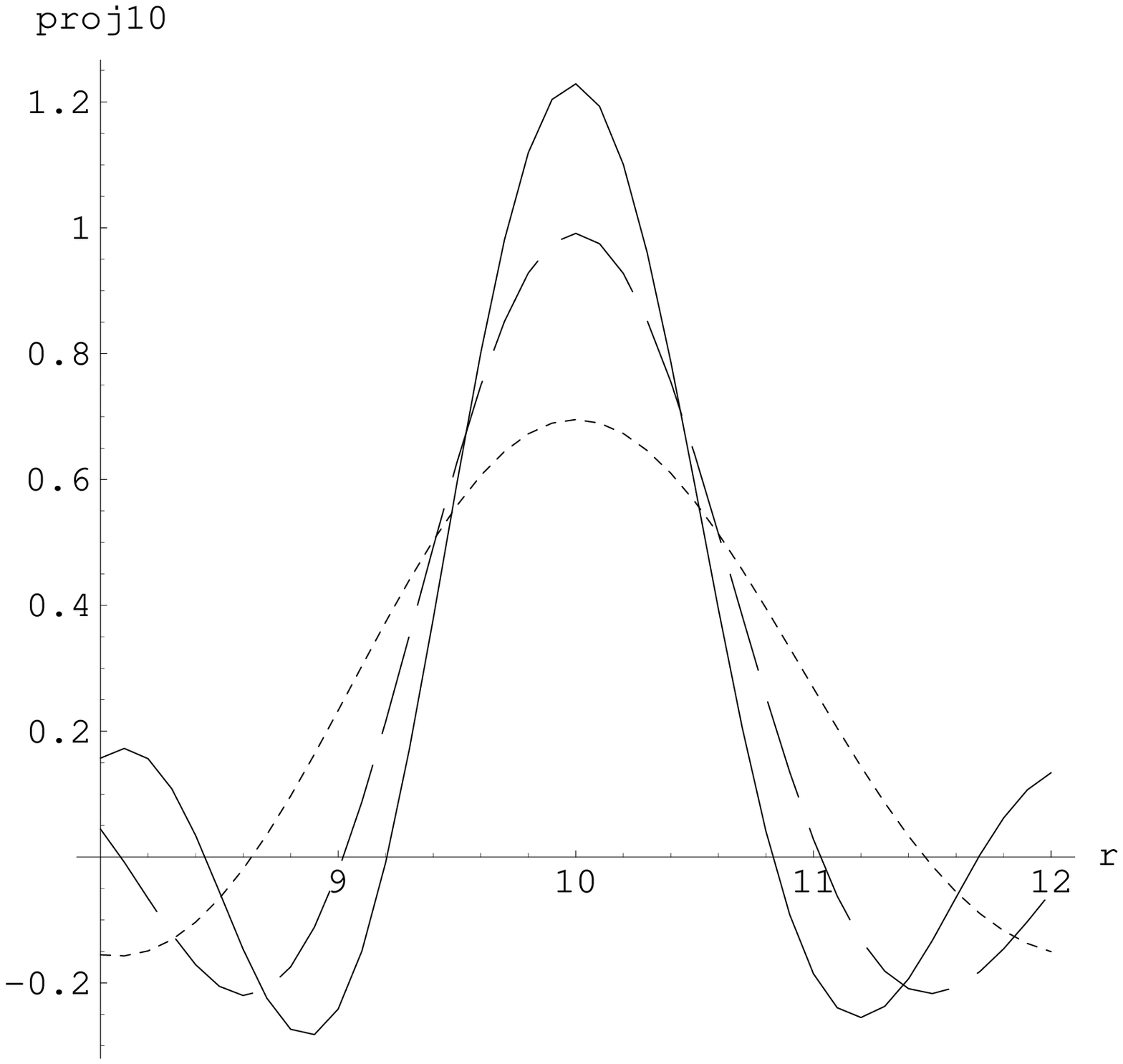}
     }
\caption{Shapes of projectors made of polynomials $G_n^1.$ Full line,
$\langle 10|{\cal P}_{150}|r \rangle,$ long dashes, 
$\langle 10|{\cal P}_{100}|r \rangle,$ short dashes
$\langle 10|{\cal P}_{50}|r \rangle.$}
\end{figure}

\begin{figure}[htb] \centering
\mbox{  \epsfysize=100mm
         \epsffile{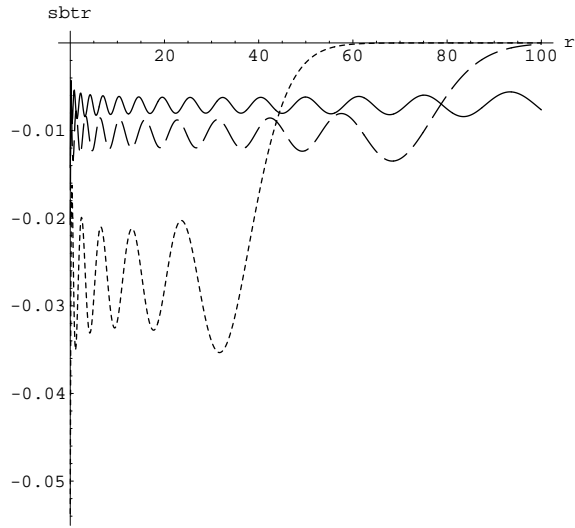}
     }
\caption{Subtractors made of $G_n^1.$ Shapes centered at $r=10.$
Short dashes, $N=10,$ long dashes, $N=20,$ full line, $N=30.$}
\end{figure}

\begin{figure}[htb] \centering
\mbox{  \epsfysize=100mm
         \epsffile{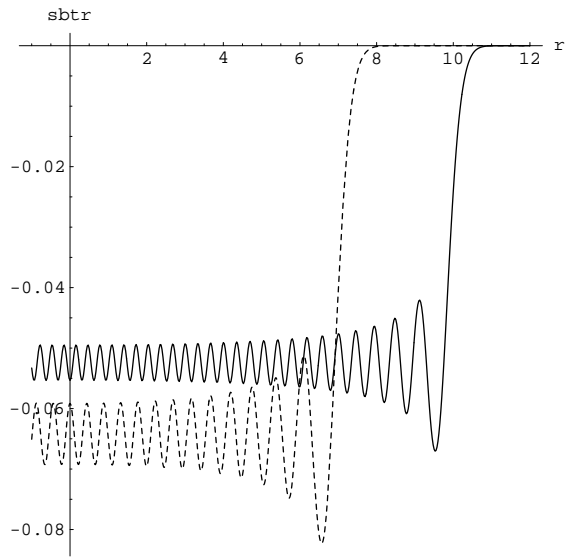}
     }
\caption{Subtractors made of $\Gamma_n.$ Shapes centered at $r=0.$ 
Stronger wiggles, shorter cut-off, dashed line, $N=50.$ Weaker wiggles, larger 
cut-off, full line, $N=100.$}
\end{figure}

\section{Illustrative example: trajectories in density space}

We return here to the toy model discussed in \cite{GiMeWe} and the 
corresponding, modified Hermite polynomials. The model consists of Z 
non interacting, spinless fermions, driven by a one dimensional harmonic 
oscillator $H_0=\frac{1}{2}(-d^2/dr^2+r^2).$ The ground state density 
from the $Z$ lowest orbitals reads, 
$\rho(r)=\sum_{i=1}^Z [\psi_i(r)]^2.$ Let $i=1,..,Z$ and $I=Z+1,...\,\infty$ 
label ``hole'' and ``particle'' orbitals, respectively. Add a perturbation 
$\delta u(r)$ to the initial potential $r^2/2.$ The first order variation of 
the density is,
\begin{equation}
\delta \rho(r)= 2 \sum_{iI} \, \psi_i(r) \psi_I(r) \,
\frac{\langle I | \delta u | i \rangle}{i-I} \, .
\label{prtrbG}
\end{equation}
If we expand $\delta u$ and $\delta \rho$ in that basis $\{w_n\}$ provided by 
the new polynomials, the formula, Eq. (\ref{prtrbG}), becomes,
\begin{equation}
\delta \rho_m= 2 \sum_{iI\, n} {\cal D}_{m\, iI} \,
\frac{ 1 } { i- I } \, {\cal D}_{n\, iI} \, \delta u_n,\ \ \ \ \ 
{\cal D}_{n\, iI} \equiv \int dr \, w_n(r) \, \psi_i(r) \psi_I(r),
\label{expansionG}
\end{equation}
where ${\cal D}$ denotes both a particle-hole matrix element of a potential 
perturbation and the projection of a particle-hole product of 
orbitals upon the basis $\{w_n\}.$ In \cite{GiMeWe} we briefly studied the 
eigenvalues and eigenvectors of this symmetric matrix,
${\cal F}={\cal D}\, (E_0-H_0)^{-1}\, \tilde {\cal D},$
where $(E_0-H_0)^{-1}$ is a short notation to account for the denominators 
and the particle-hole summation, and the tilde indicates transposition. It 
is clear that the invertible ${\cal F}$ represents the functional derivative 
$\delta \rho_m / \delta u_n$ and is suited for {\it infinitesimal} 
perturbations. We shall now take advantage of the representation provided by 
$\{ w_n \}$ to study {\it finite} trajectories $\rho(u).$ 

For this, we consider a variable Hamiltonian, 
${\cal H}_m(\lambda)=H_0+\lambda\, w_m(r),$
made of the initial harmonic oscillator, but with a finite perturbation
$\Delta u$ along one ``mode'' $w_m.$ It is trivial to diagonalize 
${\cal H}_m(\lambda)$ with an excellent numerical accuracy and thus obtain, 
given $Z,$ the ground state density $\rho(r,\lambda).$ Then it is trivial 
to expand the finite variation, $\Delta \rho=\rho(r,\lambda)-\rho(r,0),$ 
in the basis $\{w_n\}.$ This defines coordinates $\Delta \rho_n(\lambda;m)$ 
for trajectories, parametrized by the intensity of the chosen mode $m$ for 
$\Delta u.$

\begin{figure}[htb] \centering
\mbox{  \epsfysize=100mm
         \epsffile{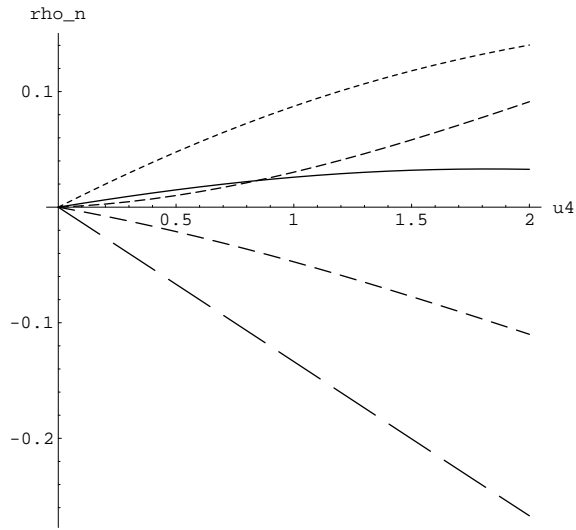}
     }
\caption{Coordinates of the perturbation density $\Delta \rho$ created by a 
perturbing potential $\Delta u=\lambda_4\, w_4.$ Full line: 
$2\, \Delta \rho_2.$ Long dashes: $\Delta \rho_4.$ Moderate dashes: 
$2\, \Delta \rho_6.$ Short dashes: $4\, \Delta \rho_8.$ Very short dashes: 
$8\, \Delta \rho_{10}.$} 
\end{figure}

\begin{figure}[htb] \centering
\mbox{  \epsfysize=100mm
         \epsffile{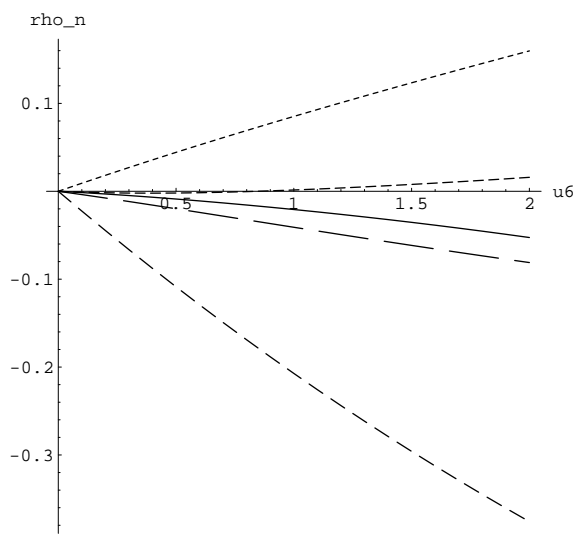}
     }
\caption{Same as Fig. 7, but now $\Delta u=\lambda_6\, w_6.$ Full line: 
$4\, \Delta \rho_2.$ Long dashes: $2\, \Delta \rho_4.$ Moderate ones: 
$\Delta \rho_6.$  Short ones: $2\, \Delta \rho_8.$ Very short
dashes: $4\, \Delta \rho_{10}.$} 
\end{figure}

\begin{figure}[htb] \centering
\mbox{  \epsfysize=100mm
         \epsffile{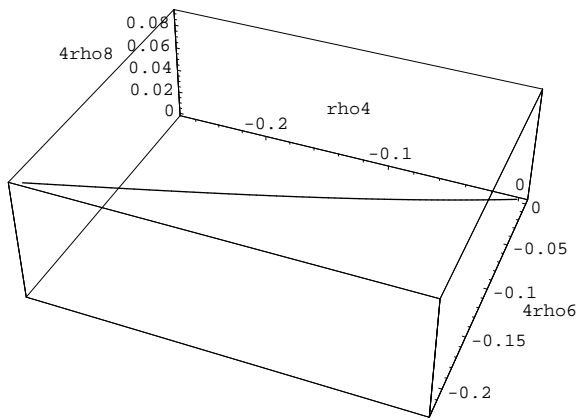}
     }
\caption{3D trajectory in density space. $\Delta \rho_4,$ $\Delta \rho_6$ and 
$\Delta \rho_8$ taken  from Fig. 7, the latter two coordinates blown $4$ 
times.} 
\end{figure}

In Figures 7 and 8 we show, with $Z=4,$ results from   
${\cal H}_4=H_0 + \lambda_4\, 2\, (2\pi)^{-\frac{1}{4}}\, 15^{-\frac{1}{2}}\, 
(8r^4-14r^2+1)\, e^{-r^2}$ and
${\cal H}_6=H_0 + \lambda_6\, (2\pi)^{-\frac{1}{4}}\, 105^{-\frac{1}{2}}\, 
(32r^6-128r^4+94r^2-11)\, e^{-r^2},$ respectively. 
The case,
${\cal H}_2=H_0 + \lambda_2\, 2\, (2/\pi)^{\frac{1}{4}}\, 3^{-\frac{1}{2}}\, 
(2r^2-1)\, e^{-r^2},$ 
makes almost a harmonic oscillator and is probably of academic interest only; 
anyhow we verified that its confirms the results with ${\cal H}_4$ and 
${\cal H}_6.$ We use a basis $\{ w_n \}$ containing a factor $e^{-r^2}$ rather
than $e^{-\frac{1}{2}r^2}$ to better match the same factor $e^{-r^2}$ created 
by products of harmonic oscillator orbitals in the calculation of matrix 
elements $\langle z_p | \Delta u | z_q \rangle,$ but this technicality is not 
important for the physics.
 
The main result to be observed seems to be the lack of ``collectivity'' for 
such modes and for such elementary Hamiltonians. Indeed, for $\lambda_4=2,$ 
the first five coordinates of $\Delta \rho$ read 
$\{0.016, -0.267, -0.055, 0.023, 0.018\},$ with a strong dominance of 
$\Delta \rho_4,$ while for $\lambda_6=2,$ these read 
$\{-0.013, -0.041, -0.376, 0.008, 0.040\},$ with a strong dominance of
$\Delta \rho_6.$  To clarify Figs. 7 and 8, we had indeed to blow up 
each $\Delta \rho_n$ by a factor $2^{|n-m|},$ where $m$ is the index of the 
driver mode in potential space. Other modes than $m=4$ and $m=6$ show the 
same property: in the density space, a trajectory driven by 
$\Delta u=\lambda w_m$ stays close to the same $w_m$ axis in that density 
space, although curvatures effects, while somewhat modest, are not absent.
Such non linearity, slight curvatures are seen in Figs. 7-8, and also in 
Figure 9, where the three $\Delta \rho_4, \Delta \rho_6, \Delta \rho_8$
sets of data shown by Fig. 7 are converted into a parametric plot for a 
trajectory. For graphical purposes again, $ \Delta \rho_6$ and $\Delta \rho_8$
are blown up $4$ times to create Fig. 9. It can be concluded, temporarily,
that the ``flexibility'' matrix ${\cal F}$ is not too far from being 
diagonal in the $\{ w_n \}$ basis, or in other words, that the $w_n$ modes 
indicate an approximately natural hierarchy in both the potential and the 
density spaces.

A subsidiary question pops up: that of the positivity of $\rho.$ Indeed, 
while the space of potentials is basically a linear space, with arbitrary 
signs for $u(r)$ when the position $r$ changes, densities $\rho(r)$ must 
remain positive for every $r.$ This creates severe constraints for any 
linear parametrization of $\Delta \rho$ in terms of the basis  $\{ w_n \}.$
In our toy model, it turns out that 
$\rho(r,0)=\pi^{-\frac{1}{2}} (8r^6-12r^4+18r+9)\, e^{-r^2}\, /6.$ Hence, if
we truncate $\Delta \rho$ to have two components only, $w_2$ and $w_4$ for 
instance, then $\rho$ is the product of $e^{-r^2}$ and a polynomial 
${\cal P}(r),$
\begin{equation}
6\, \pi^{\frac{1}{2}}\, {\cal P}(r) = 8r^6-12r^4+18r^2+9 + 
\Delta \rho_2\, 12\, (2 \pi)^{\frac{1}{4}}\, 3^{-\frac{1}{2}}\, (2r^2-1) +
\Delta \rho_4\, 12\, (\pi/2)^{\frac{1}{4}}\, 15^{-\frac{1}{2}}\, 
(8r^4-14r^2+1).
\end{equation}
Rescale out inessential factors, for a simpler polynomial, 
$\bar {\cal P}=8r^6-12r^4+18r^2+9 + \Delta R_2 (2r^2-1) + \Delta R_4 
(8r^4-14r^2+1).$
Eliminate $r$ between $\bar {\cal P}$ and $d \bar {\cal P}/dr.$ 
The resultant ${\cal R}(\Delta R_2,\Delta R_4),$ when it vanishes, 
gives the border of the convex domain of parameters $\Delta R_2, \Delta R_4$ 
where  $\bar {\cal P}$ remains positive definite. This domain contains the 
origin, because of $\rho(r,0).$ The precise form of ${\cal R}$ is 
a little cumbersome and does not need to be published here. But the resulting 
border is shown in Figure 10. Generalizations to more $\Delta \rho$ 
parameters are obvious, with more cumbersome resultants ${\cal R}.$

\begin{figure}[htb] \centering
\mbox{  \epsfysize=100mm
         \epsffile{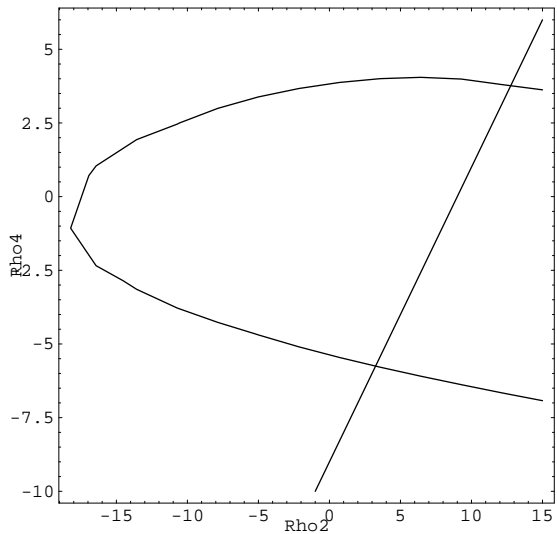}
     }
\caption{Domain of values of $\Delta R_2$ and $\Delta R_4$ acceptable
for the positivity of the density of the toy model. The domain sits 
inside the full line curve and left of the straight line. It contains the 
origin.} 
\end{figure}

\section{Discussion and Conclusion}

The subject of orthogonal polynomials has been so treated and overtreated that
any claim to novelty must contain much more than a change of the integration 
measure. We took therefore a different approach, motivated by a law of physics
and/or chemistry, matter conservation. This means a constraint of a vanishing 
average for the states described by weighted polynomials. 

For a support $[0,\infty[$ and a simple exponential weight such as 
$e^{-\frac{1}{2}r},$ a non trivial generalization of Laguerre polynomials 
occurs. This extends the generalization of Hermite polynomials described 
in \cite{GiMeWe} with the support $]-\infty,\infty[$ and Gaussian weights 
such as $e^{-\frac{1}{2}r^2}.$

We also took care of cylindrical and spherical geometries, by replacing 
$\int dr$ with $\int dr\, r$ and $\int dr\, r^2,$ respectively. The new sets
of constrained polynomials are clearly sensitive to the geometry.

For finite supports such as $[0,1]$ and constant weights, the constraint is 
already satisfied by the usual brand of orthogonal polynomials as soon as 
their order is $\ge 1.$ In that sense, we did not find significantly original 
generalizations of Legendre polynomials, although we generated polynomials 
fitted to the cylindrical and spherical geometries. The cause of the failure 
is transparent: when the weight $\mu(r)$ is a constant, there is no difference 
between the orthogonality metric $\mu^2$ and the constraint weight $\mu.$

For each set of new polynomials we found a recursion relation and a 
differential equation. There seems to be a systematic property for those
cases where the constraint generates truly original polynomials, namely 
when $\mu^2 \ne \mu.$ In such cases, recursion and differentiation seem to be 
necessarily entangled. This does not happen for traditional orthogonal
polynomials, indeed, and this ``entanglement'' may deserve some future 
attention.

Constrained polynomials expressing matter conservation in centrifuges do make
an original set if the fluid under centrifugation is compressible; a non 
constant reference weight $\mu$ is indeed in order there. But the set depends 
on the precise form of $\mu$ via potentially many physical parameters. We 
found it difficult to design, through scaling, a sufficiently ``universal'' 
set. ``Centrifuge polynomials'' will have to be calculated specifically
for each practical situation.

For those new polynomials generalizing the Hermite and Laguerre ones, we
found a description of the subspace accounting for their defect of 
completeness. A codimension 1 is the consequence of the constraint,
expressed at first by the obvious lack of a polynomial of order $n=0.$

Finally the use of such polynomials was illustrated by a toy model for
the Hohenberg-Kohn functional. A slightly surprising result was found: our
polynomials, those of low order at least, define potential perturbations 
which are reflected by density perturbations having almost the same shapes. 
This occurs despite the delocalization created by the kinetic energy operator, 
hints at short ranges in effective interactions and validates the 
localization spirit of the Thomas-Fermi method. Whether such hints are
good when the full zoology of the density functional is investigated
is, obviously, an open question; for a review of the richness of the 
functional, we refer to\cite{PK}. If long range forces are active, a 
significant amount of delocalization between the ``potential cause'' and the 
``density effect'' is not excluded. It would be interesting indeed to discover 
collective degrees of freedom in this connection between potential and density.
In any case, our main conclusion may be that the new polynomials provide, for 
the context of matter conservation, a discrete and full set of modes and 
coordinates, hence a systematic and constructive representation of phenomena.

\bigskip
It is a pleasure to thank Y. Abe, J.-P. Boujot, B. Eynard, C. Normand, 
R. Peschanski and A. Weiguny for stimulating discussions.

\end{document}